\documentstyle[12pt]{article}
\begin{document}
\begin{flushright}
LNF-96/021(IR)
\end{flushright}
\begin{flushright}
3 Maggio 96
\end{flushright}

\begin{flushleft}

{\large\bf Lower scaling dimensions of quarks and gluons and new energy scales}
\\
%\vspace{1cm}
\end{flushleft}
F. Palumbo 
\footnote{This work is carried out in the framework of the
European Community Research Program"Gauge theories, 
applied supersymmetry and quantum gravity" with a financial 
contribution under contract SC1-CT92-0789 .}
%\vspace{1cm}
\\
\par\noindent
INFN-Laboratori Nazionali di Frascati, P.O.Box 13, I-00044 Frascati, Italy

\par\noindent
\vspace{1cm}
\begin{center}
\begin{minipage}{15cm}
\small
{\bf Abstract We consider the possibility that quarks and gluons, due to
confinement, have lower scaling dimensions. In such a case there appear naturally
new energy scales below which the standard theory is recovered. Arguments are given
whereby for dimension $1/2$ of the quarks the theory is unitary also above these
energy scales.}

\end{minipage}
\end{center}
\vspace{1cm}
 
%\maketitle
\baselineskip=22pt 

1-It is generally assumed that local gauge invariance, renormalizability and 
unitarity completely determine a
lagrangian apart from mass parameters. This is certainly true for the leptons,
for which unitarity requires scaling dimension $3/2$, gauge invariance determines 
the
form of the interaction and renormalizability limits the possible terms. But for the
quarks the situation might be different. Since unitarity is needed only in the
space of physical states, the scaling dimension $3/2$ can be neither sufficient
nor necessary. Of course it is not sufficient in a perturbative framework, and
unitarity of QCD is not really proven, even though there are convincing
indications of confinement. But just because of confinement it is not obvious that
it is necessary, and one can wonder whether the quarks can have lower dimensions. 
The
same argument applies to gluons. It is the purpose  of this paper to point out that
if the scaling dimensions are lower

i) there appear naturally new energy scales below which the Standard Model is
recovered

ii) it is plausible that the dimension $1/2$ for the quarks provides the proper
scaling dimensions of hadrons also at higher energies

iii) the  dimension zero for the gluons can give a linear quark-quark potential
already at the classical level.

With lower dimensions an action renormalizabile by power counting will include terms
up to dimension 4, and it will therefore contain composite fields with
their kinetic and interaction terms. Such a 
model reduces, at low energy, to the Standard Model with the addition of
nonrenormalizable interactions suppressed by inverse powers of the new energy scales.
In order to investigate the actual dimensions of quarks and gluons one should then
examine the effects of these interactions, which might be interesting to do in
relation to the possible discrepancies betwen some recent experimental results [1]
and the Standard Model, especially in view of the fact that such discrepancies occur
only in the quark sector. But the
internal consistency of a model with lower scaling dimensions has a wider interest. 
First, if
different dimensions are theoretically possible, the ones which are not realized
 must
be forbidden by some mechanism. The exclusion of lower dimensions for the quarks 
could
be guaranteed, for instance, by a quark-lepton symmetry and would therefore be in
strong support of a Grand Unification. Note that quarks and leptons are in any case
related by the requirement of anomaly cancellation, but this does not affect the
 power
counting of the interactions [2].

Second, the problem we are considering can be formulated in a different way, 
relevant
to theories where there occur higher dimension terms . Given a Lagrangian with 
fermionic contact interactions, is it possible to make it renormalizable and
unitary? If the answer is affirmative, we can incorporate the first corrections 
in the above mentioned models in the standard framework. As far as renormalizability
by power counting is concerned, we can achieve it by lowering the dimensions of the
fundamental fermions by the addition of higher derivative terms. The real difficulty
is with unitarity.
 We will argue that this difficulty can be overcome if the elementary
fermions are confined. We will in fact show that, under certain conditions, if the
quarks have scaling dimensions $1/2$, the hadrons have the appropriate scaling
behaviour to lowest order in a perturbative scheme we are investigating.

2-Let us first define the model and show how it behaves at low energies, starting
from pure QCD. The lagrangian density of quarks of dimension $1/2$ including terms 
of dimension not greater than 4 is

\begin{eqnarray}
{\cal L}_Q &= & \sum_f -{\bar {\lambda}}_f [ {1 \over 2 } \{  D_{1f},
\rlap/{\cal{D}}\}+ \Lambda_{1f}D_{2f}+ \Lambda_f^2 (\rlap/{\cal{D}}+m_f) 
]\lambda_f    \nonumber\\
   &  &+\sum_{colorless\; composite\; fields} \alpha_M \phi(- \Box +m^2_M)\phi
+\alpha_B{\bar{\psi}}(\rlap/\partial +m_B)\psi +\alpha_Y {\bar{\psi}}\psi \phi ...
\nonumber\\
 & &+\sum_{colored\;composite \; fields}.... 
\end{eqnarray}
where
\begin{eqnarray}
{\cal{D}}_{\mu} &=&\partial_{\mu}-ig G_{\mu},
\nonumber\\
D_{if}& = &{\cal{D}}_{\mu} {\cal{D}}_{\mu}+{ 1\over 2} \gamma_{if}[ {\cal{D}}_{\mu},
{\cal{D}}_{\nu}] \sigma_{\mu \nu},\;\;\;i=1,2.
\end{eqnarray}
 
In the above equation $G_{\mu}$ is the gluon field, the m's are 
mass
parameters, the $\alpha$'s and $\gamma$'s dimensionless constants, the $\Lambda$'s
  energy scales which appear naturally due to the dimension of the quark fields
$\lambda_f$ ( of flavor $f$ ), and $\psi$ and
$\phi$  colorless composite fields with the quantum numbers
 of barions and mesons respectively. Examples of these fields, which we will need
later, are [3]

\begin{eqnarray}
\psi_{ps} &=& \epsilon_{abc}(u^a \sigma_2 \sigma_hu^b)
   (\sigma_h d^c)_s,\;\;\;for\;the\;proton
\nonumber\\
\psi_{ns} &=& \psi_{pn}(u \leftrightarrow d),\;\;\; for\;the\; neutron
\nonumber\\
\phi^+ &=&d^*{\bar{u}}^*-{\bar{d}}u ,\;\;for\;the\;\pi^+,
\end{eqnarray}
where we have adopted the convention of summation over repeated indices, a,b,c are
color indices, $u$ and ${\bar{u}}^*$ are the upper and lower components of 
$\lambda_1$
\begin{equation}
\lambda^a_{1,s}=u^a_s,\;\lambda^a_{1,s+2}=({\bar{u}}^a_s)^*,\;s=1,2,
\end{equation}
and similarly for the other quarks.  In addition there are  
in the lagrangian terms with  colored composite fields and their covariant 
derivatives. We will comment later on this proliferation of couplings.

 The free quark propagator in such a model 

\begin{equation}
{1 \over {p^2 \rlap/p  +\Lambda_{1f} p^2+ \Lambda_f^2(\rlap/p+m_f)}}
\end{equation}
becomes the usual one for $p^2 \ll \Lambda_f^2,\Lambda_{1f}\leq m_f $. We can
then consider the standard theory as a low energy approximation of the present model.
Below the energy scales $\Lambda_f$, we can introduce the
dimension $3/2$ quark field $q_f$

\begin{equation}
q_f(x)=\Lambda_f\lambda_f(x)
\end{equation}
and rescale the composites according to

\begin{equation}
\psi'=\Lambda^3 \beta_{\psi}^{-{1 \over 2}}\psi,\;\;\;\phi'=\Lambda^2
\beta_{\phi}^{ -{1 \over 2}} \phi,
\end{equation}
where $\Lambda$ is the smallest of the $\Lambda_f$'s and the $\beta$'s are
ratios of the different $\Lambda_f$, depending on the flavor content of the hadrons.
We can thus
 rewrite the quark lagrangian in the form

\begin{eqnarray}
{\cal L}'_Q &=&\sum_f {\bar {q}}_f [(\rlap/{\cal D}+m_f) +{ \Lambda_{1f} \over 
{\Lambda_f^2}} D_{2f}+ 
{1 \over {\Lambda_f^2}}{1 \over 2}\{
D_{1f}, \rlap/{\cal{D}} \} ]q_f
\nonumber\\
     & &\sum_{colorless\;composites\;fields} {\alpha_M \beta_{\phi}\over
{\Lambda^4}}\phi'(- \Box +m_M)\phi'+ {\alpha_B \beta_{\psi} \over
{\Lambda^6}}{\bar{\psi}}'(\rlap/\partial+m_B)\psi'
   +{\alpha_Y \beta_{\psi}\beta_{\phi}^{1 \over 2} \over {\Lambda^8}}{\bar
{\psi}}'\psi'\phi' ...
\nonumber\\
+ & &\sum_{colored\;composites\;fields}... 
\end{eqnarray}
It appears as the standard lagrangian with a partial ( because of the remaining
one loop divergencies ) regularization by higher derivatives. But we want to keep
$\Lambda$ finite, so that ${\cal{L}}'_Q$ is a non renormalizable lagrangian with
cutoff $\Lambda$. If the $\Lambda_f$'s are all of the same order of magnitude, the
dominant corrections come from the terms quadratic in the quark fields.

Let us now cosider the quark sector of the Standard Model. Since, as we have seen,
terms involving the composite fields are suppressed at low energy by higher inverse
powers of the energy scales, we consider here only the terms quadratic in the
left and right quark fields

\begin{equation}
{\cal{L}}_Q={\bar{\lambda}}_{L\alpha} [ \rlap/{\cal{D}} +{1 \over
{\Lambda_{L\alpha}^2}} {1 \over 2} \{ \rlap/{\cal{D}},D_{L\alpha}  \}]
\lambda_{L\alpha}+ {\bar{\lambda}}_{R\alpha} [ \rlap/{\cal{D}} +{ 1\over
{\Lambda_{R\alpha}^2}}{1 \over 2} \{ \rlap/{\cal{D}},D_{R\alpha}\}]\lambda_{R\alpha} 
\end{equation}
where $\alpha$ is the family index and, in standard notation

\begin{eqnarray}
{\cal{D}}_{\mu} &=& \partial_{\mu}-{i \over 2}g \tau W_{\mu} - {i \over 2} g'B_{\mu},
\nonumber\\
D_{L,R\alpha} &=& {\cal{D}}_{\mu} {\cal{D}}_{\mu}+\gamma_{L,R\alpha}
 [{\cal{D}}_{\mu}, {\cal{D}}_{\nu}] \sigma_{\mu\nu}.
\end{eqnarray}

 In a process involving one intermediate vector
boson, for instance, we have corrections to the SM given by terms of the type

\begin{equation}
{1 \over { \Lambda_L^2}}{\bar{\lambda}}_L {1 \over 2} \{ \Box,\rlap/{\cal{D}} \}
\lambda_L. 
\end{equation}
These are analogous to the corrections arising from the "leptophobic" 
vector boson
introduced by Altarelli et al. [4]. If we want this correction
to be smaller  than $1 \% $, we must take $\Lambda > 1 Tev$.

We conclude by  briefly considering the possibility of lower dimensions for
the gluons. In this case the scaling dimension zero is natural in the sense that it
can provide a linear potential among quarks already at the classical level [5]. The
lagrangian density including terms of dimension not greater than 4 is

\begin{equation}
{\cal {L}}_G={\cal{D}}_{\lambda}F_{\mu\nu}
{\cal{D}}^{\lambda}F^{\mu\nu} +\alpha_{G1}[F_{\mu\nu}
F^{\mu\nu}]^2  +\alpha_{G2} \Box F_{\mu\nu} F_{\mu\nu}+\Lambda_G^2
F_{\mu\nu}F^{\mu\nu}.
\end{equation}

In the above equation the $\alpha_G$'s are dimensionless constants,
$\Lambda_G$ is an energy scale, $F_{\mu\nu}$ the stress tensor and ${\cal{D}}_{\mu}$
the covariant derivative in the adjoint representation. Again one can rescale the
gauge fields and the gauge coupling ( which has here the dimension of an energy )
to check that at low energy the model reduces to the standard one  plus
nonrenormalizable interactions.

4-We now show why it is plausible that the scaling dimension $1/2$ for the
quarks is compatible with the right scaling dimensions for the hadrons and therefore
with  unitarity ( from a perturbative point of wiew) in the hadron sector. We shall
use an approach proposed [6] for the standard theory.

Consider first a barion-barion correlation function

\begin{equation}
<\psi_1(x_1)\psi_2(x_2)>={1 \over {Z_0}}\int [d{\bar{\lambda}}d \lambda]
\psi_1(x_1) \psi_2(x_2) exp[-\int d^4x {\cal {L}}_Q].
\end{equation}
We want to show that if we retain only the kinetic term for the composite field
$\psi$ in ${\cal{L}}_Q$, under certain conditions we get the right free propagator. 
The other terms must then be treated as a perturbation in the way outlined below.

 Our argument is based on the use of the composite fields as independent variables.
This requires a definition of the integral of a function of the $\psi$'s  such  that
its value be equal to that obtained by expressing the $\psi$'s in terms of the
$\lambda$'s and performing the Berezin integral over the latters [6]
\begin{equation}
\int[d\psi]f(\psi)= \int[d\lambda]f[\psi(\lambda)].
\end{equation}
Since the most general function of the $\psi$'s is a polynomial, it is sufficient
to give this definition for monomials. We say that a monomial $\Theta$ is fundamental
if it is a product of the $\psi$'s with coefficient unity and such that
\begin{equation}
\Theta_m=\psi_1^{m_1} \psi_2^{m_2}...=w_m \prod \lambda,\;\;\;w_m\neq 0.
\end{equation}
 In the above equation $\prod \lambda$ is the product of all the quark field
components in a given space-time point, $m$ is a vector index with components
$m_i=0,1$ and $w_m$ is the weight of $\Theta_m$. The definition we are looking for is
\begin{equation} \int[d\psi]\Theta_m=w_m.
\end{equation}
The important property of this definition is that if the fields $\psi$ are chosen in
such a way that there is only one fundamental monomial ( this is one of the
conditions referred to above ), the integral defined by Eq.(16) is identical, after 
a
rescaling of the $\psi$'s to get rid of the weight, to the Berezin integral. It
follows that if we assume as free action for these $\psi$-fields the Dirac action,
their propagator is the canonical propagator of a Dirac particle. Let us report, as 
an
example, that if we cofine ourselves to the quarks $u$ and $d$ the condition of a
unique fundamental monomial is naturally realized by assuming as barion variables 
the
nucleon ones. In this case in fact the only monomial with  nonvanishing weight is [6]

\begin{equation}
\Theta=\psi_{p1}\psi_{p2}\psi_{n1}\psi_{n2}=2^7 \cdot 3^2 \cdot 5 
P(u_1)P(u_2)P(d_1)P(d)2), \end{equation}
where the $\psi_{pk},\psi_{nk}$ are given by Eq.(3) and
\begin{equation}
P(u_k)=u_k^1 u_k^2 u_k^3.
\end{equation}

 Let us now consider the case of scalars. Again we can define fundamental monomials
and their integrals, but now the exponents of the bilinear composites in these
monomials can be higher than 1, and even when they are one the evaluation of the
propagators is much more complicated, and cannot in general  be performed
analytically. It has however been shown [6] that the propagator of a composite
complex scalar like $\phi$ of Eq.(3) is equal to the selfavoiding random walk in any
number of space-time dimensions where this is a generalized free theory. As it is
well known this happens in dimension greater than 4 and it is conjectured and
numerically confirmed [7] to be true also in dimension 4. Now a generalized free
field is characterized by the fact that its $n$-point functions factorize into
products of the 2-point function, but we do not know anything about the Lehmann
representation of this 2-point function, and at this stage we must assume that it is
not too different from that of a Klein-Gordon
particle. The validity of this assumption is another of the conditions referred to at
the beginning, which makes our
result for scalar composites weaker than for the spinor ones.

Processes involving at the same time nucleons and mesons can be treated for instance
by a regularization on an interpenetrating hypercubic body centered lattice, by
performing alternatively the transformation to trilinear and bilinear fields, and for
different barions and mesons we can proceed similarly.

The above results can be substantiated only if one can show in this
framework that the interactions do not violate the unitarity of the free
approximation. To do this we are studying a perturbative scheme
along the following lines. Selected the physical fields relevant in a given process,
we assume their kinetic term as the free action and treat all the other terms,
including the ones quadratic in the quark fields, as a perturbation. We then expand
wr to this perturbation and, at each order, we express everything in terms of the
fundamental monomials and perform the integration according to the above rules
assuming the relevant composites as independent variables.

 The fact that one can
construct generalized free fields by polynomials of free fields has been known for
a long time [8], but we think we have gone a little bit further. First, we have shown
that barions are canonical fields, which is encouraging in view of their behaviour
as the constituents of matter. Second, mesons are to some extent characterized and
third, the way these results are obtained opens the perspective of actual
perturbative or numerical computations.

A last comment about the proliferation of couplings. The evaluation of composite
correlation functions in the standard theory requires in general additional
normalization conditions. The additional parameters appearing with lower dimensions
might be, to some extent, the counterpart of these additional conditions. The
hadronic masses, however, in the standard theory can be calculated in terms of
the fundamental parameters, while they appear as free parameters in the present 
model. To asses whether this is really so, one should understund how the bound state
problem can be formulated in the present context.

If instead too many of the parameteres appearing in the actions (1),(9) turn out to
 be
independent and the hadronic masses cannot be calculated, even if internally
consistent the model is really unsatisfactory and it
will not be realized in Nature. The point we want to raise is that even so, its
internal consistency would have interesting consequences, pointing toward the
mechanism which prevents its occurrence.

\end{document}